\documentclass[aps,pra,showpacs,floatfix,superscriptaddress,twocolumn]{revtex4}
\usepackage[dvips]{graphicx}

\usepackage{longtable}
\usepackage{dcolumn}
\usepackage[dvips]{graphicx}
\usepackage{bm}
\usepackage{bbm}

\usepackage{times}
\usepackage{nicefrac}
\usepackage{amsmath}
\usepackage{amsfonts}
\usepackage{amssymb}
\usepackage{amsthm}

\newcolumntype{.}{D{x}{}{-1}}

\newcommand{\balpha}{\bm{\alpha}}

\newcommand{\bNabla}{\bm{\nabla}}

\newcommand{\vare}{\varepsilon}
\newcommand{\bfr}{{\bm {r}}}

\newcommand{\bfp}{{\bm {p}}}

\newcommand{\hr}{\hat{\bfr}}

\newcommand{\lbr}{\langle}
\newcommand{\rbr}{\rangle}

\newcommand{\pr}{^{\prime}}

\begin{document}

\title{Relativistic
configuration-interaction calculation of energy levels of core-excited states
in lithium-like ions: argon through krypton}

\author{V. A. Yerokhin}
\affiliation{Institute of Physics, Heidelberg University,
Im Neuenheimer Feld 226, D-69120 Heidelberg, Germany}
\affiliation{GSI Helmholtzzentrum f\"ur Schwerionenforschung, Planckstra{\ss}e 1,
D-64291 Darmstadt, Germany}
\affiliation{Center for Advanced Studies, St.~Petersburg State
Polytechnical University, Polytekhnicheskaya 29,
St.~Petersburg 195251, Russia}

\author{A. Surzhykov}
\affiliation{Institute of Physics, Heidelberg University,
Im Neuenheimer Feld 226, D-69120 Heidelberg, Germany}
\affiliation{GSI Helmholtzzentrum f\"ur Schwerionenforschung, Planckstra{\ss}e 1,
D-64291 Darmstadt, Germany}

\begin{abstract}

Large-scale relativistic configuration-interaction calculation of 
energy levels of core-excited states of lithium-like ions is presented.
Quantum electrodynamic, nuclear
recoil, and frequency-dependent Breit corrections are included in the
calculation. 
The approach is consistently applied for calculating
all $n=2$ core-excited states for all lithium-like ions starting from argon
($Z = 18$) and ending with krypton ($Z = 36$). 
The results obtained
are supplemented with systematical estimations of
calculation errors and omitted effects.

\end{abstract}

\pacs{31.30.jf, 32.30.-r, 31.15.am, 31.15.vj}

\maketitle

\section{Introduction}

Spectroscopy of lithium-like ions has received remarkable attention
over the last two decades. To a large extent,
this attention was triggered by the famous
experiment by Schweppe et al.~\cite{schweppe:91}, which demonstrated that
the $1s^22p$--$1s^22s$ energy difference in lithium-like ions can be measured with an accuracy
of better than 0.1\% of the QED contribution. This result, recapitulated and
surpassed in the following experiments
(\cite{beiersdorfer:98,brandau:04,beiersdorfer:05,epp:07,lestinsky:08}, to name a few),
made lithium-like ions arguably the best playground for testing QED
effects in the region of strong nuclear binding field.

A large number of sophisticated {\em ab initio} QED calculations have been
accomplished during the past years in order to advance theory up to the
experimental level of accuracy. In particular, the two-electron
self-energy, two-electron vacuum-polarization, and the two-photon exchange
corrections were calculated
\cite{yerokhin:99:sescr,artemyev:99,yerokhin:00:prl,sapirstein:01:lamb,yerokhin:06:prl,yerokhin:07:lilike,kozhedub:10}.
As a result, the $1s^22p_J$-$1s^22s$ transition energies in lithium-like ions are
presently among the most precise in the medium- and high-$Z$ region.
This is not the case, however, for the higher excited states of lithium-like
ions, which received much less attention from theorists. In the present
work, we aim to correct this drawback.

The object of our present study are the energy levels of the $n=2$
core-excited states of medium-$Z$ lithium-like ions. Their precise knowledge 
is required for interpretation of astrophysical spectra,
as these states contribute to the most prominent K-shell emission lines, observable
in spectra of nearly all classes of cosmic
X-ray sources (see, e.g., review~\cite{fabian:91}).
Another important motivation for studying the K-shell emission
\cite{beiersdorfer:93:aj,widmann:95} is
that it is used for the diagnostics of hot laboratory plasmas, particularly
those produced in magnetically-confined fusion research.

Accurate theoretical description of transitions involving the
core-excited states is complicated by two main issues. The first one is a
large contribution of the QED effects. Indeed, the QED
effects are strongest for the K-shell electrons and, therefore, for
the K-shell transitions. The second problem is that
the core-excited states are mostly autoionizing states. This means that
one might expect a strong mixing of the reference state with the
closely-lying continuum of single excited states (i.e., a closed core + an
electron in continuum). This interaction with continuum is very difficult to
treat accurately in theoretical calculations.

First accurate theoretical data on the core-excited states of lithium-like
ions were obtained by Vainstein and Safronova within the $1/Z$-expansion
method \cite{vainstein:78}. Later, these states were addressed
within the multiconfigurational Dirac-Fock approach
\cite{chen:86,nilsen:88,dong:06} and, more recently, within the relativistic
many-body perturbation theory \cite{safronova:04:cjp}. In the present work, we
employ the configuration-interaction (CI) method to obtain accurate values
for the Dirac-Coulomb-Breit energies, including the nuclear recoil contribution
and the frequency-dependent Breit correction. The CI energies are
supplemented with the QED correction calculated separately in the one-electron
approximation with a local screening potential. This approach is much simpler
(and, consequently, less accurate) than the full-scale QED calculations reported in
Refs.~\cite{yerokhin:00:prl,sapirstein:01:lamb,yerokhin:06:prl,yerokhin:07:lilike,kozhedub:10}
for the $1s^22p_J$-$1s^22s$ transitions, but it reproduces the results of those more complete
calculations remarkably well. Our present approach is similar to the one applied previously
by Chen et al.~\cite{chen:95} for calculating the lowest lying $n=2$ energy levels
of lithium-like ions.

The developed approach is applied for a systematic calculation of energy levels of
all $n=2$ core-excited states for all lithium-like ions starting from argon
($Z = 18$) and ending with krypton ($Z = 36$). For some of these ions, no
data on the core-excited states were previously available in the literature.

An important feature of the present investigation is a systematic
estimation of uncertainties of the calculated results. For each
particular state and each nuclear charge, we perform our CI calculations with a
large number (typically, 20-30) of different sets of configuration-state
functions. By analyzing the successive increments of the results obtained with
the set of configuration-state functions, which is increased in all possible
directions, we obtain a realistic estimate of how well our CI results have
converged. Beside the uncertainty of the Dirac-Coulomb-Breit energy, we also
estimate the uncertainty due to the omitted higher-order QED effects. This is done by
performing the QED calculations with three different screening potentials
and analyzing the dependence of the results on the choice of the potential.

The remaining paper is organized as follows. In the following section, we give a
brief outline of our implementation of the CI method.
The evaluation of the QED correction is discussed
in Sec.~\ref{sec:QED}. Sec.~\ref{sec:num} presents the numerical details of
the calculations.
Finally, in Sec.~\ref{sec:res}, we report the results
of our calculations and compare them with data available in the literature.
Relativistic units $\hbar=c=1$ and charge units $e^2/4\pi  = \alpha$
are used throughout this paper.


\section{Configuration interaction method}
\label{sec:CI}

\subsection{DCB Hamiltonian}

Relativistic Dirac-Coulomb-Breit (DCB) Hamiltonian of an $N$-electron atom with an
infinitely heavy nucleus is given by
\begin{equation}\label{eq1}
    H_{\rm DCB} = \sum_i h_{\rm D}(i) + \sum_{i<j} \left[ V_{C}(i,j)+
    V_{B}(i,j)\right]\,,
\end{equation}
where indices $i,j = 1,\ldots,N$ numerate the electrons, $h_D$ is the one-particle
Dirac Hamiltonian,
\begin{equation}\label{eq2}
    h_D(i) = \balpha_i\cdot\bfp_i+ (\beta-1)\,m+ V_{\rm nuc}(r_i)\,,
\end{equation}
$\balpha$ and $\beta$ are the Dirac matrices, $\bfp$ is the momentum operator,
$V_{\rm nuc}$ is the binding
potential of the nucleus, $V_{C}$ and $V_B$ are the Coulomb and the
(frequency-independent) Breit parts of the electron-electron interaction,
\begin{align}\label{eq3a}
&V_{C}(i,j) = \frac{\alpha}{r_{ij}}\,, \\
\label{eq3}
V_{B}(i,j) =  &\ -\frac{\alpha}{2\,r_{ij}}\,
    \left[ \balpha_i\cdot\balpha_j + \left( \balpha_i\cdot \hr_{ij}\right)
             \left( \balpha_j\cdot \hr_{ij}\right) \right]\,,
\end{align}
with $r_{ij} = |\bm{r}_i-\bm{r}_j|$ and $\hr = \bfr/r$. The operator
$H_{\rm DCB}$ acts in the space of the
positive-energy eigenfunctions of some one-particle Dirac Hamiltonian
$\tilde{h}_D$, which does not necessarily coincides with
(\ref{eq2}). It is usually convenient to include a part of the
electron-electron interaction effect already in the one-electron orbitals, by
introducing a screening potential $V_{\rm scr}(r)$ and defining $\tilde{h}_D$ as
\begin{equation}\label{eq4a}
    \tilde{h}_D(i) = \balpha_i\cdot\bfp_i+ (\beta-1)\,m+ V_{\rm nuc}(r_i) + V_{\rm scr}(r_i)\,.
\end{equation}

The $N$-electron wave function of the system with parity $P$, angular momentum
quantum number $J$, and its projection $M$ is represented as a linear combination
of configuration-state functions (CSFs),
\begin{equation}\label{eq4}
  \Psi(PJM) = \sum_{r=1}^{N_{\rm CSF}} c_r \Phi(\gamma_r PJM)\,,
\end{equation}
where $\gamma_r$ denotes the set of additional quantum numbers that determine the
CSF. The CSFs are constructed as antisymmetrized products of one-electron orbitals
$\psi_n$ of the form
\begin{equation}\label{eq5}
    \psi_n(\bfr) = \frac1r\,         \left(
        \begin{array}{c}
        G_n(r)\,\chi_{\kappa_n m_n}(\hr)\\
      i F_n(r)\,\chi_{-\kappa_n m_n}(\hr)   \\
        \end{array}
        \right)\,,
\end{equation}
where $\chi_{\kappa m}$ the spin-angular spinor \cite{rose:61}, $\kappa =
(-1)^{j+l+1/2}(j+1/2)$ is the relativistic angular parameter, and $m$ is the
angular momentum projection. In the present work, we chose the one-electron
orbitals $\psi_n$ to be the (positive-energy) eigenfunctions of the one-electron
Dirac Hamiltonian (\ref{eq4a}) with the screening potential being
the frozen-core Dirac-Fock potential $V_{DF}$, defined by by its action on a wave
function as
\begin{align}\label{eq5b}
    V_{DF}(\bfr_1)\,\psi(\bfr_1) &\, = \sum_c \int d\bfr_2\,
      \psi_c^+(\bfr_2)\,\frac{\alpha}{r_{12}}\,
 \nonumber \\ &\, \times
        \Bigl[\psi_c(\bfr_2)\,\psi(\bfr_1)-\psi_c(\bfr_1)\,\psi(\bfr_2)\Bigr]\,,
\end{align}
where the index $c$ runs over the core orbitals.

In the CI method, the energy levels of the system and the mixing coefficients
$c_r$ in Eq.~(\ref{eq4}) are obtained by solving the secular equation
\begin{equation}\label{eq6}
    {\rm det} \bigl\{\lbr \gamma_r PJM|H_{\rm DCB}|\gamma_s PJM\rbr -E_r\,\delta_{rs}\bigr\} =
    0\,.
\end{equation}
The matrix elements of the Hamiltonian between the CSFs can be represented as
linear combinations of the one- and two-particle radial integrals \cite{johnson:book},
\begin{align}\label{eq7}
\lbr \gamma_r PJM| &\,  H_{\rm DCB}|\gamma_s PJM\rbr = \sum_{ab}
 d_{rs}(ab)\,I(ab)
  \nonumber \\
 + &\, \alpha \sum_k \sum_{abcd} v_{rs}^{(k)}(abcd)\,
  R_{k}(abcd)\,.
\end{align}
Here, $a$, $b$, $c$, and $d$ specify the one-electron orbitals, $d_{rs}$ and
$v^{(k)}_{rs}$ are the angular coefficients, $I(ab)$ are the one-electron radial
integrals, and $R_{k}(abcd)$ are the relativistic generalization
of Slater radial integrals. Further details of our implementation of the CI
method can be found in our previous study \cite{yerokhin:08:pra}.

\subsection{Corrections to DCB energies}

In the present work, we include two corrections to the DCB Hamiltonian
(\ref{eq1}), namely the nuclear recoil and the frequency-dependent Breit
corrections. The nuclear recoil correction is small and will be treated
nonrelativistically. It is conveniently
separated into two parts, the normal mass shift (NMS) and the specific mass
shift (SMS). The NMS part can be easily factorized out and
accounted for by multiplying the eigenvalues of the DCB Hamiltonian
(\ref{eq1}) by the reduced mass prefactor
\begin{equation}
 E_{\rm DCB} \to \,\frac{\mu}{m}\,E_{\rm DCB}\,,
\end{equation}
where $\mu/m = 1/(1 + m/M)$ and
$m$ and $M$ are masses of the electron and the nucleus,
respectively. The SMS part of the recoil effect is accounted for by adding
an additional term to the DCB Hamiltonian,
\begin{align}\label{4eq3}
  \delta H_{\rm SMS} = \frac{m}{M}\,\sum_{i<j} \bfp_i\cdot\bfp_j\,.
\end{align}

The frequency-dependent Breit correction comes from the QED theory. It can be
obtained by substituting the Coulomb $V_C$ and the standard Breit $V_B$
interactions by the full QED electron-electron interaction operator,
when calculating the matrix elements of the DCB Hamiltonian
(\ref{eq1}) with the reference-state wave function(s). Note that in our
work, contrary to the approach sometimes used in the literature (e.g.,
in Ref.~\cite{chen:95}), we do not construct the
whole DCB Hamiltonian matrix with the frequency-dependent Breit interaction,
but apply the frequency-dependent Breit interaction for the reference CSFs
only. The reason is that the DCB Hamiltonian (\ref{eq1}) was derived
\cite{sucher:80} for the standard (frequency-independent) Breit interaction
only, and inclusion of the frequency dependency for highly excited CSFs can
lead to sizeable spurious effects.

The QED electron-electron interaction operator
$I(\omega)$ depends on the energy of the virtual photon $\omega$ and has different forms
in different gauges. In the Feynman gauge, it is given by
\begin{equation}
 I_{\rm Feyn}(\omega,r_{ij}) =   \alpha\,\bigl( 1-\balpha_i \cdot \balpha_j \bigr)\,
  \frac{e^{i\,|\omega|\,r_{ij}}}{r_{ij}} \,,
\end{equation}
whereas in the Coulomb gauge, it reads
\begin{align}
 I_{\rm Coul}(\omega,r_{ij}) &= I_{\rm Feyn}(\omega,r_{ij})
  \nonumber \\ &
-
  \alpha\, \left[ 1+\frac{\bigl(\balpha_i \cdot \bNabla_i\bigr)\bigl(\balpha_j\cdot \bNabla_j\bigr)}{\omega^2} \right]\,
  \frac{e^{i\,|\omega|\,r_{ij}}-1}{r_{ij}} \,.
\end{align}
General QED expressions for the one-photon exchange correction for a given
reference state \cite{shabaev:02:rep}
involve two kind of matrix elements, the
direct ones of the form $\lbr ab| I |ab \rbr$ and the exchange
ones of the form $\lbr ba| I |ab \rbr$. The energy of the virtual photon is
fixed by $\omega = 0$ in the direct terms and by $\omega = \vare_a-\vare_b$ in
the exchange terms. In the case of several equivalent reference
CSFs, there are also non-diagonal terms and consequently, matrix elements
of a general form $\lbr ab| I |cd \rbr$. For them, we used an energy
symmetrized expression \cite{artemyev:05:pra}
\begin{equation}
\lbr ab| I |cd \rbr \to
\frac12\left[\lbr ab| I(\vare_{a}-\vare_c) |cd \rbr+\lbr ab|
  I(\vare_b-\vare_d) |cd \rbr\right]\,.
\end{equation}
It can be easily verified that, for a single reference state, the matrix
elements of the operator $I$ are gauge independent. In the case of several reference-state
configurations, gauge invariance is not exact, but we checked that the
residual is completely negligible in each particular case.


\section{Radiative QED corrections}
\label{sec:QED}

In this section, we deal with corrections to the DCB energies due to
QED effects. For the purpose of the present investigation, it will be
sufficient to take into account only the dominant part of the radiative QED
effects, given by the one-electron self-energy and the vacuum polarization
calculated in a local screening potential.

The total QED
contribution for a given many-electron reference state is obtained by summing
the QED contributions from the one-electron orbitals, weighted by their
fractional occupation numbers as obtained from the eigenvectors of the CI
calculation,
\begin{align}
\delta E_{\rm QED} = \sum_a q_a\,\bigl[\lbr a| \Sigma_{\rm SE}(\vare_a)|a\rbr
+  \lbr a| V_{\rm VP} |a\rbr  \bigr]\,,
\end{align}
where index $a$ runs over all one-electron orbitals contributing to the given
many-electron state, $q_a$ is the occupation number of the one-electron
orbital, $\Sigma_{\rm SE}$ is the self-energy operator, $\vare_a$ is the Dirac
energy of the one-electron state $a$, and $V_{\rm VP}$ is the vacuum
polarization potential.

The one-electron self-energy correction in a general local potential is calculated
rigorously, to all orders in the nuclear binding strength parameter
$Z\alpha$. The method used in the present work is quite close to the one
developed by Blundell and Snyderman \cite{blundell:91:se}. We, however,
introduced several improvements to their original calculation
scheme. First, we used the dually kinetically balanced $B$-spline basis set
\cite{shabaev:04:DKB}, which yielded significantly better convergence
than the original $B$-spline method by
Johnson et al.~\cite{johnson:88}. Second, we evaluated the many-potential
electron propagator $G^{(2+)}$ by forming the difference
\begin{align}
G^{(2+)} = G - G^{(0)} - G^{(1)}\,,
\end{align}
where $G$, $G^{(0)}$, and $G^{(1)}$ are the full electron propagator, the free
propagator, and the one-potential propagator, respectively. The one-potential
propagator was calculated by taking a numerical derivative of the $B$-spline basis
set with respect to the nuclear charge $Z$,
\begin{align}
G^{(1)} = Z\, \left( \frac{d}{dZ}\,G \right)_{Z = 0}\,.
\end{align}
We found that this approach
yields results equivalent to those obtained by the direct (triple) summation over the
spectrum but is much faster in actual computations.

The one-electron vacuum-polarization contribution was straightforwardly
calculated as an expectation value of the Uehling and Wichmann-Kroll
potentials with the one-electron orbitals. The Wichmann-Kroll potential was
evaluated by the approximate formulas tabulated in Ref.~\cite{fainshtein:91}.

In our calculations of the one-electron self-energy and vacuum-polarization
corrections, we assume that the initial state and all intermediate states in
the electron propagator are the eigenstates of
the Dirac Hamiltonian (\ref{eq4a}) with a screening potential $V_{\rm scr}$.
The screening potential allows us to account for the dominant part of the
electron-electron interaction and should be chosen with care.
It would be natural to use the same screening potential as in the CI part of
our calculations, but this is not possible since the Dirac-Fock potential
(\ref{eq5b}) is non-local. Because of
this, we introduce three different
local screening potentials, whose eigenstates and eigenvalues are
quite close to the Dirack-Fock ones.

Each potential is constructed into two steps.
First, we solve the Dirac-Fock problem (for the
center-of-gravity of all equivalent relativistic configurations) and find the
Dirac-Fock wave functions. Second, we use these wave function to set up
our local screening potentials.

The simplest choice is the potential induced by the charge density
of all other electron orbitals except the reference one. This is the
core-Hartree (CH) potential defined, for a given one-electron state $a$, as
\begin{equation}\label{eqq4}
    V_{{\rm CH},a}(r) = \alpha \int_0^{\infty} dr\pr \frac1{\max(r,r\pr)}\, \rho_a(r\pr)\,,
\end{equation}
where $\rho_a$ is the density of all one-electron
orbitals excluding $a$,
\begin{equation}\label{eeq5}
    \rho_a(r) = \sum_{b\neq a} q_b\, \left[ G_b^2(r)+F_b^2(r)\right]\,,
\end{equation}
with $G_b$ and $F_b$ being the upper and the lower components of the Dirac-Fock
solution. Note that the CH potential defined in this way is different from the
one frequently
encountered in the literature because we do not require self-consistency
for $V_{{\rm CH},a}$.

The second choice of the screening potential is a variant of the Kohn-Sham
potential \cite{kohn:65,sapirstein:02:lamb}, defined as
\begin{eqnarray}\label{eeq8}
    V_{\rm KS}(r) &=& \alpha \int_0^{\infty} dr\pr \frac1{\max(r,r\pr)}\,
        \rho(r\pr)
 \nonumber \\ &&
        - \frac23\, \frac{\alpha}{r}\,
          \left[ \frac{81}{32 \pi^2}\,r\,\rho(r) \right]^{1/3}
      - \frac{\alpha}{r}\,\left[1- e^{-(A r)^2} \right]
          \,,\nonumber \\
\end{eqnarray}
where $\rho(r)$ is the {\em total} Dirac-Fock charge density
\begin{equation}
    \rho(r) = \sum_{b} q_b\, \left[ G_b^2(r)+F_b^2(r)\right]\,.
\end{equation}
The second term in the right-hand side Eq.~(\ref{eeq8}) is the exchange
correction derived from density functional theory, whereas the third term is a
kind of the Latter correction \cite{latter:55}, which restores the physical
asymptotical behaviour of the KS potential at large distances. The parameter
$A$ in the Latter correction was chosen to be about $Z\alpha/10$. Again, the
KS potential (\ref{eeq8}) is different from the one encountered in the
literature  because we do not require self-consistency in its definition.

The third screening potential employed in this work is referred to as the Localized
Dirac-Fock (LDF) potential. For a given state $a$, the LDF potential $V_{{\rm
    LDF},a }$ is obtained by inverting the Dirac Fock equations with the known
solutions $G_a$ and $F_a$ and then smoothing out the result in the vicinity of
zeros of $G_a$. The LDF potential was first introduced for calculations of
the QED corrections to the parity-nonconserving transition amplitudes in caesium
\cite{shabaev:05:prl,shabaev:05:pra}. In the present work, we employ the
variant of the LDF potential described in detail in Ref.~\cite{yerokhin:07:lilike}.

According to our experience and comparison with the results of more complete
calculations, the LDF potential yields better results in
calculations of the QED effects than the KS and CH potentials. Therefore, we
will use this potential for obtaining the final results for the QED
correction in our present calculation.

%
%
%
\begin{table}
\caption{Contributions to the Coulomb energy (in a.u.)
for the $1s2s2p\, ^4P^o_{1/2}$ state of lithium-like iron ($Z=26$), for the
infinitely heavy nucleus. The values listed after the first row are the
increments obtained on successively adding configurations while increasing
the maximal value of the orbital quantum number $L_{\rm max}$
and enlarging the size of the one-electron basis.
Label $SD$ stands for the single and double excitations,
$T$ denotes the triple excitations, $n_a$ is the number of $B$-splines in the
one-electron basis set, $\epsilon$ is energy
the cutoff parameter, see text for the details.
\label{tab:CI:C}}
\begin{ruledtabular}
  \begin{tabular}{l.}
  $L_{\rm max}$ & \delta E   \\
    \hline\\[-5pt]
    \multicolumn{2}{l}{$SD$, $n_a = 30$, $\epsilon = 4$}\\
   1  &   -497.762x\,342\,48  \\
   2  &     -0.008x\,719\,04  \\
   3  &     -0.000x\,662\,72    \\
   4  &     -0.000x\,124\,13    \\
   5  &     -0.000x\,034\,81    \\
   6  &     -0.000x\,012\,45    \\
   7  &     -0.000x\,005\,18    \\
$\ge 8$  &     -0.000x\,005\,55    \\[3pt]
    \multicolumn{2}{l}{$SD$, $n_a = 40$, $\epsilon = 8$}\\
   1  &     -0.000x\,173\,23    \\
   2  &     -0.000x\,001\,34    \\
   3  &     -0.000x\,001\,13    \\
$\ge 4$  &     -0.000x\,001\,07    \\[3pt]
    \multicolumn{2}{l}{$SD$, $n_a = 50$, $\epsilon = 16$}\\
   1  &     -0.000x\,006\,46    \\
   2  &     -0.000x\,000\,43    \\[3pt]
    \multicolumn{2}{l}{$T$, $n_a = 25$, $\epsilon = 4$}\\
   1  &     -0.000x\,000\,22    \\
   2  &     -0.000x\,000\,77    \\
   3  &     -0.000x\,000\,09    \\
$\ge 4$  &     -0.000x\,000\,04    \\[3pt]
Final result   &   {\bf -497.772}x\,{\bf 091}\,{\bf (7)} \\
  \end{tabular}
\end{ruledtabular}
\end{table}

%
%
%
\begin{table}
\caption{Contributions to the frequency-independent
Breit energy (in a.u.)
for the $1s2s2p\, ^4P^o_{1/2}$ state of lithium-like iron ($Z=26$), for the
infinitely heavy nucleus. Notations are the same as in Table~\ref{tab:CI:C}.
\label{tab:CI:B}}
\begin{ruledtabular}
  \begin{tabular}{l.}
  $L_{\rm max}$ & \delta E   \\
    \hline\\[-5pt]
    \multicolumn{2}{l}{$SD$, $n_a = 30$, $\epsilon = 4$}\\
   1   &     0.065x\,471\,70    \\
   2   &    -0.000x\,330\,01    \\
   3   &    -0.000x\,050\,23    \\
   4   &    -0.000x\,014\,47    \\
   5   &    -0.000x\,005\,98    \\
$\ge 6$   &    -0.000x\,009\,15    \\[3pt]
    \multicolumn{2}{l}{$SD$, $n_a = 40$, $\epsilon = 8$}\\
   1   &    -0.000x\,001\,76    \\
   2   &    -0.000x\,000\,43    \\
   3   &    -0.000x\,000\,25    \\[3pt]
    \multicolumn{2}{l}{$SD$, $n_a = 50$, $\epsilon = 16$}\\
   1   &     0.000x\,004\,49    \\[3pt]
Final result    &     {\bf 0.065}x\,{\bf 064}\,{\bf (9)} \\
  \end{tabular}
\end{ruledtabular}
\end{table}

%
%
%
\begin{table*}
\caption{QED contributions (in a.u.)
for the $1s2s2p\, ^4P^o_{1/2} - 1s^22s\, ^2S_{1/2}$ transition in lithium-like iron
($Z=26$), for different screening potentials. $q$ denotes the
occupation number of the single electron orbitals as obtained from the CI calculations.
SE, VP, and QED label the
self-energy, the vacuum-polarization, and the total QED contribution, respectively.
\label{tab:qed}}
\begin{ruledtabular}
  \begin{tabular}{lc...c....}
 & \multicolumn{4}{c}{$^4P^o_{1/2}$} & \multicolumn{4}{c}{$^2S_{1/2}$}
    & \multicolumn{1}{c}{$^4P^o_{1/2}-^2S_{1/2}$}\\[3pt]
State & $q$ & {\rm SE} & {\rm VP} & {\rm QED} & $q$ & {\rm SE} & {\rm VP} & {\rm
  QED} &  \\
\hline
\\[-5pt]
\multicolumn{4}{l}{LDF potential:}\\
$1s$ &        1.000  & 0.15x688 & -0.01x337 &  0.14x350  & 2.000 & 0.30x432 & -0.02x610 & 0.27x822 & -0.13x472 \\
$2s$ &        1.000  & 0.01x974 & -0.00x158 &  0.01x816  & 1.000 & 0.01x894 & -0.00x153 & 0.01x741 & 0.00x075 \\
$2p_{1/2}$ &   0.943 &-0.00x052 & -0.00x001 &  -0.00x053  & & & & & -0.00x053 \\
$2p_{3/2}$ &   0.057 & 0.00x005 &  0.00x000 &   0.00x005  & & & & & 0.00x005 \\
Sum &      &         &         &          &            & & & & -0.13x445 \\[3pt]
\multicolumn{4}{l}{KS potential:}\\
$1s$ &        1.000  & 0.15x512 &  -0.01x326  & 0.14x186 & 2.000 & 0.30x287 &-0.02x600 & 0.27x687 & -0.13x501 \\
$2s$ &        1.000  & 0.01x991 &  -0.00x159  & 0.01x832 & 1.000 & 0.01x922 &-0.00x154 & 0.01x768 & 0.00x064 \\
$2p_{1/2}$ &   0.943 &-0.00x053 &  -0.00x001  &-0.00x054 & & & & & -0.00x054 \\
$2p_{3/2}$ &   0.057 & 0.00x005 &   0.00x000 &  0.00x005 & & & & & 0.00x005 \\
Sum  &     &         &         &          &            & & & & -0.13x487 \\[3pt]
\multicolumn{4}{l}{CH potential:}\\
$1s$ &        1.000  & 0.15x674 &  -0.01x337 &   0.14x338 & 2.000 & 0.30x384 & -0.02x607 & 0.27x778 & -0.13x440 \\
$2s$ &        1.000  & 0.01x956 &  -0.00x157 &   0.01x799 & 1.000 & 0.01x894 & -0.00x153 & 0.01x741 & 0.00x058 \\
$2p_{1/2}$ &   0.943 &-0.00x051 &  -0.00x001 &  -0.00x052 & & & & & -0.00x052 \\
$2p_{3/2}$ &   0.057 &0.00x005  &   0.00x000 &   0.00x005 & & & & & 0.00x005 \\
Sum   &    &         &         &          &            & & & & -0.13x429 \\[3pt]
Final result   &    &         &         &          &            & & & & {\bf
  -0.13}x{\bf 445}\,{\bf (43)} \\[3pt]
  \end{tabular}
\end{ruledtabular}
\end{table*}

\section{Numerical details}
\label{sec:num}

In the present work, we use our implementation of the CI method employed
previously in Refs.~\cite{yerokhin:08:pra:ra,yerokhin:08:pra} for the
evaluation of the hyperfine splitting in Li and Be$^+$. The one-electron basis
was constructed with help of the dual-kinetic-balance $B$-spline
method \cite{shabaev:04:DKB}. The screening potential in the one-electron
Hamiltonian (\ref{eq4a}) was taken to be the frozen-core DF potential
(\ref{eq5b}), with the $1s^2$ core in the case of the ground and
valence-excited states and the $1s$ core in the case of the core-excited
states. The standard Fermi model of the nuclear charge distribution was used
to represent the nuclear potential.

For a given number of B-splines $n_a$, all Dirac eigenstates were taken with the energy
$0< \vare_n \le mc^2(1+Z\alpha\, \epsilon)$ and the orbital quantum number $l \le
L_{\rm max}$, where the value of the energy cutoff parameter
$\epsilon$ was varied between $4$ and $16$ and $L_{\rm max}$, between $1$ and $7$.
In order to check the convergence of our CI results with respect to the
number of partial waves included and the size of the basis, we perform a set
of calculations with different number of basis functions and analyze the successive
increments as the size of the basis is enlarged. In our analysis, we study the
dependence of the results obtained on the parameters $n_a$, $\epsilon$,
and $L_{\rm max}$, and the type of virtual excitations included.
In all cases
relevant for the present study, the contribution of triple excitations was
found to be very small. We thus perform our calculations first with
including single and double excitations only and then adding the contribution of the triples
calculated separately with a smaller basis.
The analysis of the convergence of the partial-wave expansion is supplemented
with an estimation of the omitted tail, which is obtained by a polynomial
least-squares fitting of the increments in $1/l$.

An example of our CI
calculations is presented in Table~\ref{tab:CI:C} for the Dirac-Coulomb energy
and in Table~\ref{tab:CI:B} for the Breit correction. The results reported in
these tables are obtained by performing calculations with 27 different sets of basis
functions. Each
set is characterized by parameters $n_a$, $\epsilon$, and $L_{\rm
  max}$. E.g., for $n_a = 30$, $\epsilon = 4$, and $L_{\rm max }=7$, the
one-electron basis reads $17s\,$$16p\,$$16d\,$$15f\,$$15g\,$$15h\,$$14i\,$$14k\,$,
where $16p$ means 16 lowest-lying orbitals of the $p_{1/2}$ symmetry {\em and}
16 lowest-lying orbitals of the $p_{3/2}$ symmetry, etc.
Generating all possible
single- and double-excited CSF's with this set of one-electron orbitals leads
to a basis of about $N_{\rm CSF}=30,000$ functions. As another example, for
$n_a = 50$, $\epsilon = 16$, and $L_{\rm max}=2$, the set of one-electron
orbitals is $29s\,$$29p\,$$28d\,$, leading to the basis of about $N_{\rm CSF}=25,000$
functions.

Analysis of data presented in Tables~\ref{tab:CI:C} and \ref{tab:CI:B}
allows us to give a reliable estimate of the uncertainty of our CI
calculations. In the case shown in these tables, the dominant error comes from the omitted
tail of the partial-wave expansion. For higher core-excited states, however, the
dominant error often comes from the convergence of the one-electron basis
of $np$ and $nd$ symmetries, which is explained by strong mixing of the
reference (autoionizing, core-excited) states with the closely-lying
continuum of single-excited $1s^2np$ and $1s^2nd$ virtual states.
In some cases (particularly, for the $^2D$ states), we had to employ large
one-electron basis sets, up to $40s\,$$39p\,$$39d\,$ ($N_{\rm CSF} = 47,000$ functions),
in order to ensure the convergence of CI results.

We now discuss our calculations of the radiative QED corrections.
Table~\ref{tab:qed} presents a
detailed breakdown of individual QED contributions for the $1s2s2p\,
^4P^o_{1/2} - 1s^22s\, ^2S_{1/2}$ transition in lithium-like iron. The
calculation is performed for the LDF, KS, and CH potentials. All three results
for the transition energy agree very well with each other.
Remarkably, the agreement is much better for the energy difference than for
each energy level separately. The same situation occurs for the other
transitions from the core excited states to the ground state.

As a final result for the QED contribution, we take the value obtained with the
LDF potential (since this potential yields the best agreement with the results
of more complete calculations). The uncertainty of the QED contribution was
estimated as the maximal deviation of the KS and CH values from the LDF one.
We checked that such estimation of uncertainty is consistent with the results
of more sophisticated QED calculations available
for the $1s^22p_{J}$ states
\cite{yerokhin:99:sescr,artemyev:99,kozhedub:10}. The set of calculations
presented in Table~\ref{tab:qed} takes about 30h of the processor time on a
modern workstation.

\section{Results}
\label{sec:res}

Table~\ref{tab:main} presents our calculation results for the energy levels
of the $1s^22s$, $1s^22p$, and  $1s\,2l\,2l'$
states of lithium-like
ions, from argon ($Z = 18$) to krypton ($Z = 36$).
The total energies are listed in the table for the ground
state, whereas for all other states, relative energies with respect to the
ground state are given. Note that for the $1s^22p_J$-$1s^22s$
energy differences, much more accurate theoretical results are available in the
literature \cite{yerokhin:06:prl,kozhedub:10}. We nethertheless present our own
results for the $1s^22p_J$ states in the table since they (i) provide us with an independent check of
accuracy of our calculations and (ii) allow us to deduce the
transition energies between the core-excited and the $1s^22p_J$
states.

The third and fourth columns of Table~\ref{tab:main} present our results for the
Dirac-Coulomb and Breit energies (multiplied by the reduced mass prefactor).
For each value of
the nuclear charge $Z$ and each state, the CI calculation was performed with
20-30 different basis sets. After that, an extrapolation of the
partial-wave expansion to $L_{\rm max}=\infty$ was performed and an estimation of
the uncertainty was made by analyzing the results as illustrated by
Tables~\ref{tab:CI:C} and \ref{tab:CI:B}. The procedure for the error
estimation is fully automatized and done consistently in the same way
for all $Z$ and all states. Each set of CI calculations for one value of $Z$
and one state takes about 20-50h of processor time on a modern workstation.

The fifth column of Table~\ref{tab:main} presents our calculation results for the QED
correction. For each value of
$Z$ and each state, the calculation was performed with
three screening potentials. The results listed in the table are obtained with
the LDF potential. The maximal deviation of the KS and CH results from the LDF
one was taken as the uncertainty of the QED correction.
Small additional contributions to the energy levels from the specific mass shift and the
frequency-dependent Breit correction are listed in the sixth and seventh
columns of Table~\ref{tab:main}, respectively.

One may note that the uncertainties of our calculation results  
presented in Table~\ref{tab:main} have some
irregularities along the $Z$ sequence. For
some values of $Z$ the estimated uncertainty turns out to be much larger than
for the neighbouring ones. This is caused by a non-uniform convergence of the
results with respect to the size of the 
basis set, which is due to the interaction of the 
reference autoionizing state with the
continuum of single-excited states. In the CI calculations, the continuum is
discretized and, if a continuum state in the basis happens to be nearly
degenerate in energy with the reference state, a sizeable non-physical contribution
may arise. (It is non-physical since the true Green
function $G(\vare)$ is known to be regular at energies $\vare > m$.) When this
occurs, we observe irregularity in the convergence of our results and have to increase
the uncertainty accordingly. Note that for a given state, we use exactly the
same basis for all values of $Z$ listed in Table~\ref{tab:main} and estimate
the uncertainty exactly in the same way, without any smoothing of energies
or uncertainties along the $Z$ sequence. 

Our final theoretical results for the wavelengths of the 22 strongest $1s2l2l'\to
1s^22l'$ transitions in lithium-like ions are summarized in Table~\ref{tab:wav}. 
The transitions are labelled from ``{\em a}'' to ``{\em v}'', following the
standard notations as by Gabriel~\cite{gabriel:72}.

We now turn to the comparison of our calculation results 
for the three most important elements, namely argon, iron, and krypton, with data
previously reported in the literature. 
Since the amount of the 
experimental and theoretical data available for these ions 
is rather voluminous, we decided to restrict
ourselves to a comparison with the relativistic many-body perturbation calculation
by Safronova and Safronova \cite{safronova:04:cjp} and with the
extensive compilations of the theoretical and experimental world data 
\cite{saloman:10,shirai:00,saloman:07}. The calculation by Safronova and Safronova
is apparently the best
of the previous theoretical studies of the core-excited states and the compilations
\cite{saloman:10,shirai:00,saloman:07} are the primary sources for the NIST
recommended spectral data for these ions \cite{nist:12}. 
In the case of argon, the values reported in
the compilation by Saloman~\cite{saloman:10} include the results by
Safronova and Safronova but in addition provide estimations of uncertainties, so
we compare only with Saloman's compilation.

The comparison for argon, iron, and krypton
is presented in Tables~\ref{tab:Ar}, \ref{tab:Fe}, and \ref{tab:Kr}, respectively. In the case of argon, the
compilation \cite{saloman:10} includes estimations of uncertainties.
The agreement is nearly always within the given
error bars but our results are significantly more accurate.
For iron and krypton, there are no estimations of uncertainty given in previous
works, but the probable errors might be guessed from the differences between
different results. Agreement of our results with results by Safronova and Safronova is in
most cases better than that with the NIST compilation data, especially in the case of
krypton.

Comparison of our calculation results for other ions with the NIST spectral
data \cite{nist:12} looks similar to what can be seen in
Tables~\ref{tab:Ar}, \ref{tab:Fe}, and \ref{tab:Kr}. Root-mean-square deviation of 
energies of core-excited states varies from 0.002~Ry for argon to 0.05~Ry for
krypton and grows smoothly with the increase of the nuclear charge. The only
large deviation is found for the $1s2s2p\,^2P^o_{1/2}$ state of copper, for which
NIST database reports $615.26$~Ry whereas our result is
$612.319\,(1)$~Ry. This deviation is most probably due to a misprint in the
NIST database. No data on energies of core-excited states is available
in the database for $Z = 30$ and $Z = 32-35$ and only few levels are available for
$Z = 20$ and $21$.

In summary, we performed a large-scale relativistic CI calculation of the
energy levels of all $n = 2$ core-excited states of lithium-like ions from
argon to krypton. The CI results include the nuclear recoil contribution
and the frequency-dependent Breit correction. The
QED correction was calculated separately in the one-electron
approximation with a local screening potential.
The calculation results are supplemented with a systematic
estimation of uncertainties. The uncertainties of the CI values were evaluated
by analyzing the successive increments of the results obtained with
the set of configuration-state functions increased in all possible
directions. The uncertainty of the QED corrections were evaluated by
performing the QED calculations with three different screening potentials
and analyzing the dependence of the results on the choice of the potential.

\section*{Acknowledgement}

Stimulating discussions with Jos\'e Crespo L\'opez-Urrutia are gratefully
acknowledged.
We are grateful to Prof.~I.~Tupitsyn for providing us with his version of
CI code for checking and comparison and for many useful advices on
implementation of the CI method.
The work reported in this paper was supported by the Helmholtz Gemeinschaft
(Nachwuchsgruppe VH-NG-421).

%
%
%
\begingroup
\begin{longtable*}{llcdddddd}
\caption{Energy levels of lithium-like ions from argon through krypton, in Rydbergs,
1 Ry = 109\,737.315\,685\,39\,(55)~cm$^{-1}$. Separately listed are the
Dirac-Coulomb energy, the frequency-independent Breit correction, the QED
correction, the specific mass shift (SMS) correction, and the frequency-dependent
Breit (Breit$'$) correction. The total energies are presented for the ground state,
whereas
for all other states, the energies relative to the ground state are given.
$\mu/m$ is the reduced mass prefactor and $R$ is the nuclear root-mean-square charge radius.
\label{tab:main}}\\
\colrule\hline
\multicolumn{2}{c}{Term} & $J$ &
\multicolumn{1}{c}{$\ \ $Coulomb} &
\multicolumn{1}{c}{$\ \ \ \ \ $Breit} & \multicolumn{1}{c}{$\ \ \ \ \ $QED} &
\multicolumn{1}{c}{$\ \ \ \ \ $SMS} &
\multicolumn{1}{c}{$\ \ \ \ \ $Breit$^{\prime}$} &
\multicolumn{1}{c}{$\ \ \ \ \ $Total}
\\
\hline
\\ \multicolumn{6}{l}{
{\bf Ar  (Z =  18)},\ \  $1-\mu/m = 0.00001373$,\ \ $R = 3.427$\ fm} \\[2pt]
$1s^22s$             & $^2S$   & 1/2 &      -696.0032 &    0.1504 &    0.1676 &    0.0001 &   -0.0001 &      -695.6852 \, (4) \\
$1s^22p$             & $^2P^o$ & 1/2 &         2.3412 &    0.0113 &   -0.0098 &   -0.0006 &   -0.0003 &         2.3417 \, (6) \\
                     &         & 3/2 &         2.5867 &   -0.0015 &   -0.0092 &   -0.0006 &   -0.0004 &         2.5749 \, (8) \\
$1s2s^2$             & $^2S$   & 1/2 &       226.5266 &   -0.1362 &   -0.0669 &   -0.0001 &    0.0001 &       226.3234 \, (17) \\
\hline
\end{longtable*}
\endgroup

%
%
%
\begingroup
\squeezetable
\begin{ruledtabular}
\begin{longtable*}{cllllllllllllll}
\caption{Theoretical wavelengths of the $1s 2l2l'\to 1s^2 2l'$ transitions in
  lithlium-like ions, in \AA. Transitions are labelled as by Gabriel \cite{gabriel:72}.
\label{tab:wav}}\\
\colrule\hline\\[-6pt]
 Key & \multicolumn{1}{c}{Transition} & \multicolumn{1}{c}{$Z = 18$} & \multicolumn{1}{c}{$Z = 19$} & \multicolumn{1}{c}{$Z = 20$} & \multicolumn{1}{c}{$Z = 21$} & \multicolumn{1}{c}{$Z = 22$} & \multicolumn{1}{c}{$Z = 23$} & \multicolumn{1}{c}{$Z = 24$}\\
 \hline\\[-6pt]
$a$ & $1s(^2S)2p^2(^3P)\,^2P_{3/2} \to 1s^22p\,^2P^o_{3/2}$  &  3.985716 (12)&  3.562881 (7)&  3.203642 (6)&  2.895849 (6)&  2.630130 (5)&  2.399149 (5)&  2.197103 (4) \\
$b$ & $1s(^2S)2p^2(^3P)\,^2P_{3/2} \to 1s^22p\,^2P^o_{1/2}$  &  3.981655 (8)&  3.558756 (6)&  3.199456 (6)&  2.891608 (5)&  2.625838 (4)&  2.394811 (4)&  2.192723 (4) \\
$c$ & $1s(^2S)2p^2(^3P)\,^2P_{1/2} \to 1s^22p\,^2P^o_{3/2}$  &  3.989796 (22)&  3.567118 (7)&  3.208023 (7)&  2.900372 (6)&  2.634791 (6)&  2.403940 (5)&  2.202015 (5) \\
$d$ & $1s(^2S)2p^2(^3P)\,^2P_{1/2} \to 1s^22p\,^2P^o_{1/2}$  &  3.985727 (20)&  3.562983 (7)&  3.203826 (6)&  2.896118 (5)&  2.630484 (5)&  2.399585 (4)&  2.197616 (4) \\
\hline
\end{longtable*}
\end{ruledtabular}
\endgroup

%
%
%
\begin{table*}
\caption{Energy levels of Ar XVI ($Z=18$), relative to the ground state.
Comparison is made with the recent compilation of theoretical and experimental
world data by Saloman~\cite{saloman:10}. Units are Rydbergs.
\label{tab:Ar}}
\begin{ruledtabular}
  \begin{tabular}{lc..}
  State & $J$ &\multicolumn{1}{c}{Present work} & \multicolumn{1}{c}{Saloman \cite{saloman:10}}  \\
    \hline\\[-5pt]
$1s^22p\,^2P^o$& 1/2   & 2.x3417 \,(6) & 2.x342\,193\,(36) \\
               & 3/2   & 2.x5749 \,(8) & 2.x575\,268\,(36) \\
$1s(^2S)2s^2\,^2S$& 1/2     & 226.x3234\,(17) & 226.x324\,(12) \\
$1s(^2S)2s2p(^3P^o)\,^4P^o$& 1/2   & 226.x9032 \, (4)& 226.x915\,(17) \\
               & 3/2   & 226.x9658 \, (4)& 226.x977\,(17) \\
               & 5/2   & 227.x1215 \, (3)&  \\
$1s(^2S)2s2p(^3P^o)\,^2P^o$& 1/2   & 228.x7624 \, (9)& 228.x766\,(17) \\
               & 3/2   & 228.x8852 \, (4)& 228.x887\,(17) \\
$1s(^2S)2p^2(^3P)\,^4P$& 1/2     & 229.x4514 \, (6)& 229.x467\,(12) \\
             & 3/2     & 229.x5553 \, (6)& 229.x569\,(12) \\
             & 5/2     & 229.x6689 \,(6) & 229.x682\,(17) \\
$1s(^2S)2s2p(^1P^o)\,^2P^o$& 1/2   & 229.x6226 \, (39)& 229.x625\,(17) \\
               & 3/2   & 229.x6720 \, (43)& 229.x678\,(17) \\
$1s(^2S)2p^2(^1D)\,^2D$& 3/2     & 230.x7372 \, (33)& 230.x741\,(12) \\
             & 5/2     & 230.x7449 \, (35) & 230.x746\,(17) \\
$1s(^2S)2p^2(^3P)\,^2P$& 1/2     & 230.x9743 \, (12)& 230.x975\,(12) \\
             & 3/2     & 231.x2081 \, (7)& 231.x208\,(12) \\
$1s(^2S)2p^2(^1S)\,^2S$& 1/2     & 232.x361 \, (13)& 232.x368\,(13) \\
  \end{tabular}
\end{ruledtabular}
\end{table*}

%
%
%
\begin{table*}
\caption{Energy levels of Fe XXIV ($Z=26$), relative to the ground state.
Comparison is made with the compilation of theoretical and experimental
world data by Shirai et al.~\cite{shirai:00} and with recent calculation
by Safronova and Safronova \cite{safronova:04:cjp}.
Units are Rydbergs.
\label{tab:Fe}}
\begin{ruledtabular}
  \begin{tabular}{lc...}
    State & $J$ &\multicolumn{1}{c}{Present work} &
  \multicolumn{1}{c}{Shirai \cite{shirai:00}}  &
     \multicolumn{1}{c}{Safronova \cite{safronova:04:cjp}}  \\
    \hline\\[-5pt]
$1s^22p\,^2P^o$& 1/2   & 3.x5708 \, (15) &   3.x572\,01 &   3.x575 \\
               & 3/2   & 4.x7442 \, (16) &   4.x745\,49 &  4.x749 \\
$1s(^2S)2s^2\,^2S$& 1/2     & 485.x1587 \, (16) &   485.x120  &  485.x165 \\
$1s(^2S)2s2p(^3P^o)\,^4P^o$& 1/2   & 486.x0358 \, (8)&   486.x097  &   486.x053 \\
               & 3/2   & 486.x3082 \, (8)&   486.x316  &   486.x326 \\
               & 5/2   & 487.x1491 \, (7)&               &   487.x165 \\
$1s(^2S)2s2p(^3P^o)\,^2P^o$& 1/2   & 488.x9700 \, (18)&  489.x021   &   488.x980 \\
               & 3/2   & 489.x6618 \, (8)&   489.x644  &   489.x668 \\
$1s(^2S)2p^2(^3P)\,^4P$& 1/2     & 490.x2797 \, (15)&  490.x307  &   490.x305 \\
             & 3/2     & 490.x8937 \, (16)&  490.x918  &   490.x919 \\
             & 5/2     & 491.x3308 \, (16)&  491.x337  &   491.x358 \\
$1s(^2S)2s2p(^1P^o)\,^2P^o$& 1/2   & 490.x6865 \, (34)&  490.x709   &   490.x695 \\
               & 3/2   & 490.x9054 \, (45)&  490.x909   &   490.x915 \\
$1s(^2S)2p^2(^1D)\,^2D$& 3/2     & 492.x6885 \, (17)&  492.x638  &   492.x708 \\
             & 5/2     & 493.x1184 \, (16)&  493.x104  &   493.x130 \\
$1s(^2S)2p^2(^3P)\,^2P$& 1/2     & 492.x7648 \, (16)&  492.x777  &   492.x772 \\
             & 3/2     & 494.x1071 \, (16)&  494.x084  &   494.x112 \\
$1s(^2S)2p^2(^1S)\,^2S$& 1/2     & 495.x5424 \, (19)&  495.x489  &   495.x559 \\
  \end{tabular}
\end{ruledtabular}
\end{table*}

%
%
%
\begin{table*}
\caption{Energy levels of Kr XXXIV ($Z=36$), relative to the ground state.
Comparison is made with the compilation of theoretical and experimental
world data by Saloman~\cite{saloman:07} and with recent calculation
by Safronova and Safronova \cite{safronova:04:cjp}.
Units are Rydbergs.
\label{tab:Kr}}
\begin{ruledtabular}
  \begin{tabular}{lc...}
  State & $J$ &\multicolumn{1}{c}{Present work} &
  \multicolumn{1}{c}{Saloman \cite{saloman:07}}  &
     \multicolumn{1}{c}{Safronova \cite{safronova:04:cjp}}  \\
    \hline\\[-5pt]
$1s^22p\,^2P^o$& 1/2   &  5.x2334\,(33) &   5.x23598 & 5.x242 \\
               & 3/2   &  10.x0061 \, (36)&  10.x00839    & 10.x02 \\
$1s(^2S)2s^2\,^2S$& 1/2     & 950.x4035 \, (25) & 950.x49       & 950.x42 \\
$1s(^2S)2s2p(^3P^o)\,^4P^o$& 1/2   & 951.x7127 \, (19) & 951.x96 & 951.x75 \\
               & 3/2   & 952.x4125 \, (37) & 952.x69 & 952.x44 \\
               & 5/2   & 956.x3541 \, (15) &           & 956.x38 \\
$1s(^2S)2s2p(^3P^o)\,^2P^o$& 1/2   & 956.x069 \, (27) &  956.x33 & 956.x09 \\
               & 3/2   & 959.x5618 \, (17) & 959.x62 & 959.x58 \\
$1s(^2S)2p^2(^3P)\,^4P$& 1/2     & 958.x5153 \, (30) & 958.x88 & 958.x56 \\
             & 3/2     & 961.x9480 \, (35) & 962.x13 & 962.x00 \\
             & 5/2     & 962.x7869 \, (36) & 962.x97  & 962.x83 \\
$1s(^2S)2s2p(^1P^o)\,^2P^o$& 1/2   & 961.x104 \, (15) & 961.x21 & 961.x12 \\
               & 3/2   & 961.x719 \, (34) & 961.x80 & 961.x74 \\
$1s(^2S)2p^2(^1D)\,^2D$& 3/2     & 964.x4259 \, (36) & 964.x61 & 964.x45 \\
             & 5/2     & 967.x7716 \, (37) & 967.x76 & 967.x80 \\
$1s(^2S)2p^2(^3P)\,^2P$& 1/2     & 964.x0167 \, (34) & 964.x21 & 964.x03 \\
             & 3/2     & 969.x5580 \, (36) & 969.x52 & 969.x57 \\
$1s(^2S)2p^2(^1S)\,^2S$& 1/2     & 971.x1155 \, (57) & 971.x08 & 971.x14 \\
  \end{tabular}
\end{ruledtabular}
\end{table*}

\end{document}